\documentclass{article}
\usepackage{spconf,amsmath,graphicx}
\usepackage{xcolor}
\usepackage{enumitem}
\setlist{nosep, leftmargin=14pt}

\usepackage{mwe} 

\usepackage{booktabs}
\usepackage{bm}
\usepackage[super]{nth}
\usepackage{amssymb}
\usepackage{hyperref}
\usepackage[symbol]{footmisc}

\newif\ifcomment
\commenttrue
\ifcomment
\newcommand{\shirin}[1]{{\bf \textcolor{purple}{Shirin: #1}}}
\newcommand{\poojitha}[1]{{\bf \textcolor{orange}{Poojitha: #1}}}

\else
\newcommand{\shirin}[1]{}
\newcommand{\poojitha}[1]{}
\fi

\title{Demonstration of an Adversarial Attack Against a Multimodal Vision Language Model for Pathology Imaging}
\name{
    Poojitha Thota$^{\star, 1}$ \qquad
    Jai Prakash Veerla$^{\star, 1, 3}$ \qquad
    Partha Sai Guttikonda$^{1, 3}$ 
}

\secondlinename{
    Mohammad S. Nasr$^{1,3}$ \qquad
    Shirin Nilizadeh$^{\dagger, 1, 3}$ \qquad
    Jacob M. Luber$^{\dagger, 1, 2, 3}$
}

\address{
    $^{1}$ Department of Computer Science and Engineering, University of Texas at Arlington\\
    $^{2}$ Department of Bioengineering, University of Texas at Arlington\\
    $^{3}$ Multi-Interprofessional Center for Health Informatics, University of Texas at Arlington
}

\begin{document}
\maketitle
    \def\thefootnote{$\star$}\footnotetext{ Co-Authors - Equal contribution; Order of authors is alphabetical and both first authors have the right to list their name first on their respective CVs.} \def\thefootnote{$\dagger$}\footnotetext{Responsible authors. Email: \texttt{jacob.luber@uta.edu}, \texttt{shirin.nilizadeh@uta.edu}}
    
    \begin{abstract}
In the context of medical artificial intelligence, this study explores the vulnerabilities of the Pathology Language-Image Pretraining (PLIP) model, a Vision Language Foundation model, under targeted attacks. Leveraging the Kather Colon dataset with 7,180 H\&E images across nine tissue types, our investigation employs Projected Gradient Descent (PGD) adversarial perturbation attacks to induce misclassifications intentionally. The outcomes reveal a 100\% success rate in manipulating PLIP's predictions, underscoring its susceptibility to adversarial perturbations. The qualitative analysis of adversarial examples delves into the interpretability challenges, shedding light on nuanced changes in predictions induced by adversarial manipulations. These findings contribute crucial insights into the interpretability, domain adaptation, and trustworthiness of Vision Language Models in medical imaging. The study emphasizes the pressing need for robust defenses to ensure the reliability of AI models. The source codes for this experiment can be found at \href{https://github.com/jaiprakash1824/VLM_Adv_Attack}{https://github.com/jaiprakash1824/VLM\_Adv\_Attack}.
\end{abstract}

    \section{Introduction}
\label{sec:intro}

\begin{figure}[htb]

\begin{minipage}[b]{1.0\linewidth}
  \centering
  \centerline{\includegraphics[width=8.5cm]{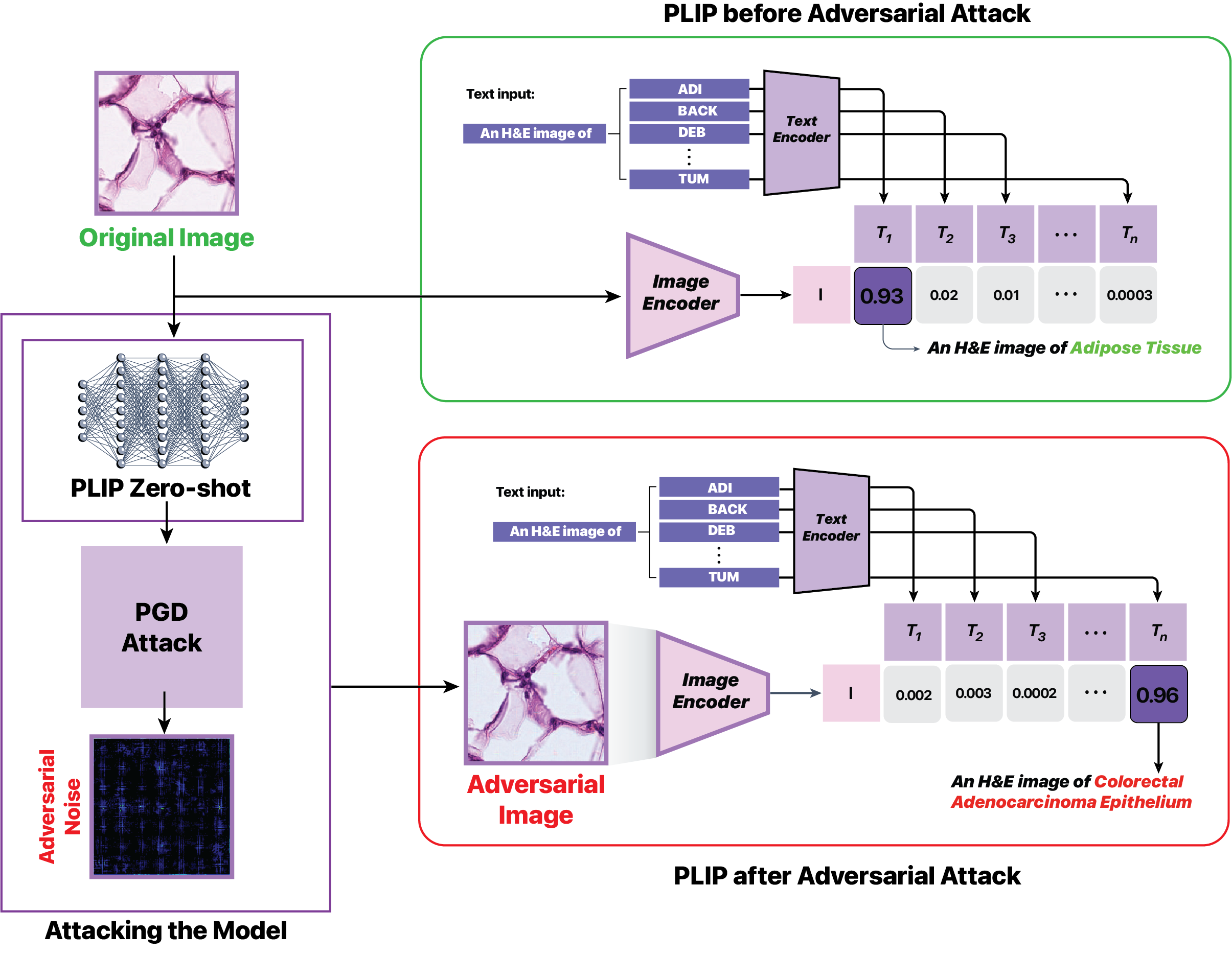}}
\end{minipage}
\caption{Attack Overview}
\label{fig:1}
\end{figure}
The incorporation of artificial intelligence (AI) into the field of medical imaging and pathology has experienced notable advancements, leading to substantial progress in the areas of diagnoses and analysis~\cite{lu2021ai,ng2023prospective}. Furthermore, the utilization of AI in the field of pathology goes beyond traditional imaging methods and encompasses the intricacies associated with Hematoxylin and Eosin (H\&E) staining~\cite{de2021deep}. This integration allows for a thorough comprehension of tissue structures and the morphology of cells.

 PLIP, a vision language model for pathology, is an AI framework that operates in several dimensions, effectively managing the complex interplay between visual and textual data~\cite{Huang2023}. It leverages the collective knowledge found on platforms such as medical Twitter to overcome the limitations posed by a scarcity of annotated medical images. This significant advancement not only tackles the issue of limited data availability but also enables pioneering research in the field of pathology analysis. These models have exhibited exceptional performance in zero-shot classification, representing cutting-edge capabilities in this domain.






\begin{figure}[t]
    \centering
    \centerline{\includegraphics[width=10cm, height=15cm, keepaspectratio]{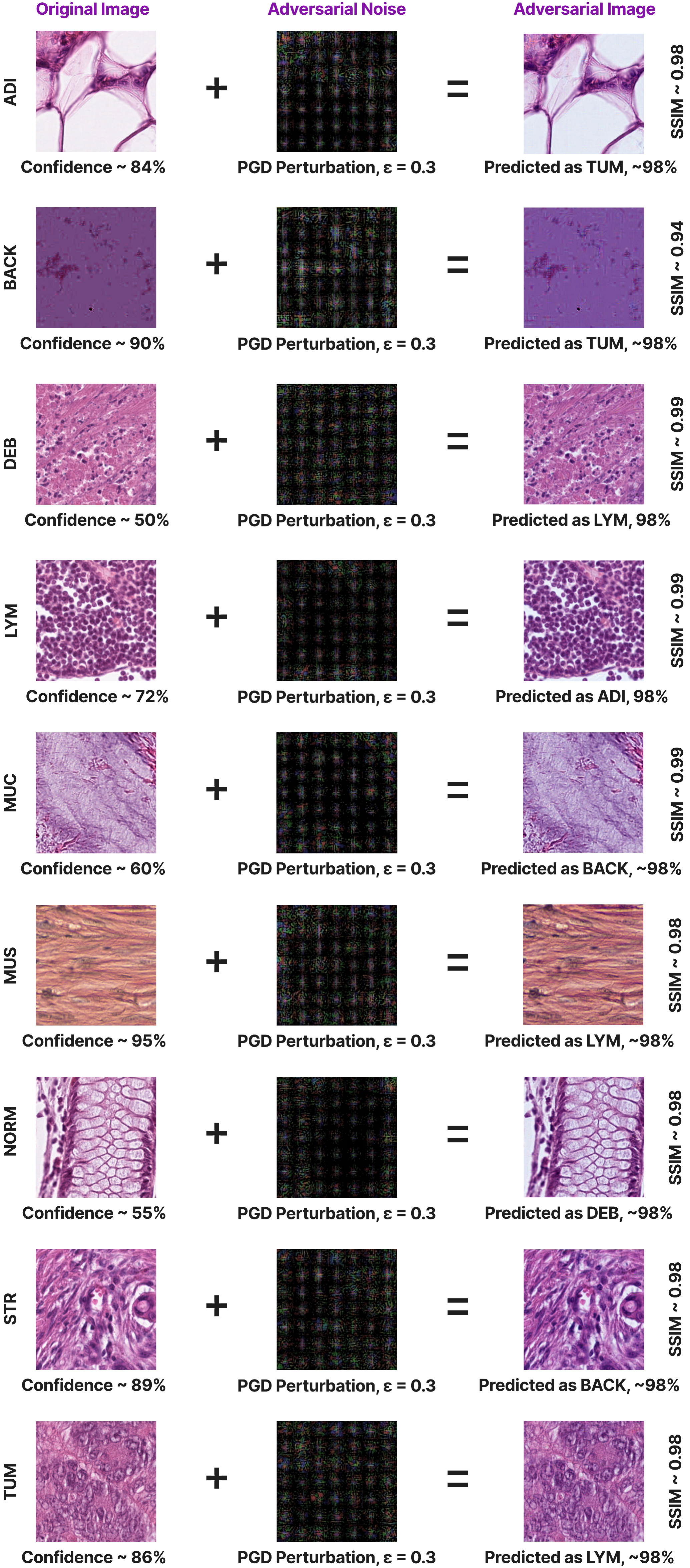}}
    \caption{Original H\&E + Pertubation = Adversarial H\&E }
    \label{fig:2}
\end{figure}
With the increasing integration of AI models like PLIP into pathology practice, the potential threat of adversarial attacks becomes a significant concern. The reliability of AI systems in medical imaging can be compromised by many techniques, such as Fast Gradient Sign Method (FGSM) attack~\cite{goodfellow2014explaining}, Carlini \& Wagner~\cite{carlini2017towards},  and projected gradient descent (PGD) attack~\cite{madry2019deep}. These attacks have the potential to pose significant challenges to the integrity and accuracy of AI systems in this domain. 

Interpretability and trustworthiness are very important to the field of pathology. The lack of transparency and comprehensibility in the decision-making process of numerous machine learning models gives rise to issues regarding their black-box character and the capacity to understand the underlying reasoning behind their forecasts. 


In this domain, the ability to make informed decisions relies heavily on the comprehensibility of the outputs generated by AI systems.  By conducting a thorough investigation of attacks and emphasizing the need for interpretability and trustworthiness, our objective is to provide significant contributions to the ongoing academic discussion regarding the crucial impact of AI on the development of pathology analysis in the future through the demonstration of the first succesful adversarial attack against a pathology vision language model. 


    \section{methods}
\label{sec:methods}


\subsection{Threat Model}
\label{ssec:Threat Model}

Our examination of the threat model centers on the reliability of Vision Language Models (VLMs), specifically PLIP. A hypothetical situation where such an adversarial attack might manifest would be a payer altering diagnostic results with the intention of obtaining financial benefits, such as denying an insurance claim. 

\subsection{Vision Language Models - PLIP}
\label{ssec:VLM-PLIP}

Our investigation focuses on the methodological aspects of the PLIP  model, including examining its architectural intricacies. PLIP has been chosen based on its exceptional ability to include both visual and textual data. The multimodality inherent to vision language models is achieved by leveraging medical Twitter to carefully select and organize the OpenPath dataset. This dataset is a substantial resource consisting of 208,414 pathology images accompanied by descriptions in natural language. This model has been rigorously evaluated on popular datasets, such as KatherColon~\cite{kather2019predicting}, PanNuke~\cite{gamper2019pannuke}, DigestPath~\cite{da2022digestpath} datasets for zero-shot image classification.     


\subsection{Dataset}
\label{ssec:Dataset}

\begin{figure}[htb]
    \centering
    \centerline{\includegraphics[width=1.0\linewidth]{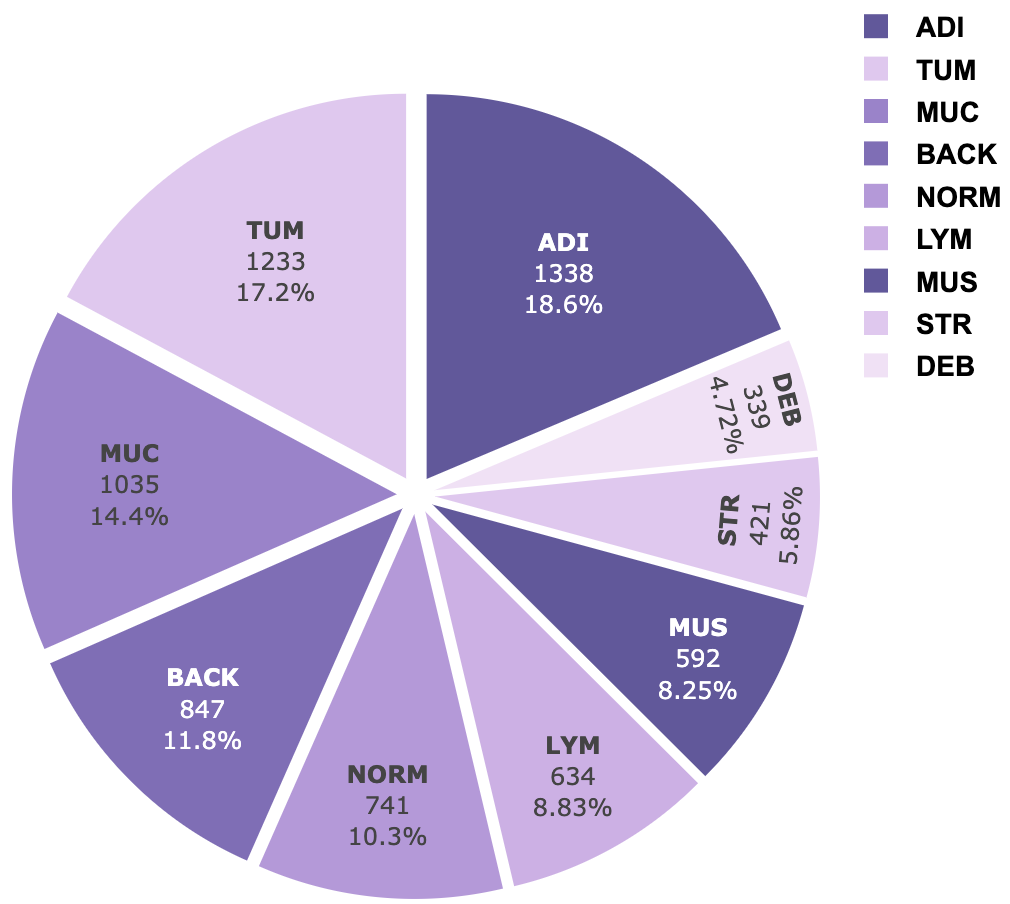}}
    \caption{PieChart of Kather Colon Dataset}
    \label{fig:3}
\end{figure}
Our study relies on the Kather Colon dataset, a carefully selected 7,180 image patches from 50 colorectal adenocarcinoma patients. 

The dataset used in this study consists of photographs with dimensions of 224x224 pixels and a resolution of 0.5 megapixels per pixel (MPP). It serves not only as a testing platform but also as a validation set with clinical relevance, hence assuring the practicality and application of our research findings in real-world scenarios. Fig.\ref{fig:3}, is a piechart representing 9 tissue types with the number of images in each tissue type.

\subsection{Adversarial Attack - Projected Gradient Descent}
\label{ssec:Adversarial Attack}


We chose to perform adversarial attack on PLIP by performing the PGD method. The choice of PGD is supported by its efficacy as an iterative optimization method, enabling the creation of perturbations that are both effective and subtle.  This particular attribute contributes to our objective of methodically examining and comprehending the vulnerabilities of PLIP under controlled hostile circumstances. This enables a thorough evaluation of the resilience of PLIP against sophisticated attacks.

As shown in Fig. \ref{fig:1}, we take the original H\&E image and perform the PGD attack by iterating through several steps to achieve optimal perturbation. We perform this attack based on two main objectives, which are targeted misclassification and a high Structural Similarity Index Measure (SSIM) score.
    \section{ Experiements and Results}
\label{sec:results}

\subsection{Evaluation}
\label{ssec:Evaluation}



\begin{figure}[t]
\centering
{\includegraphics[width=0.4\textwidth]{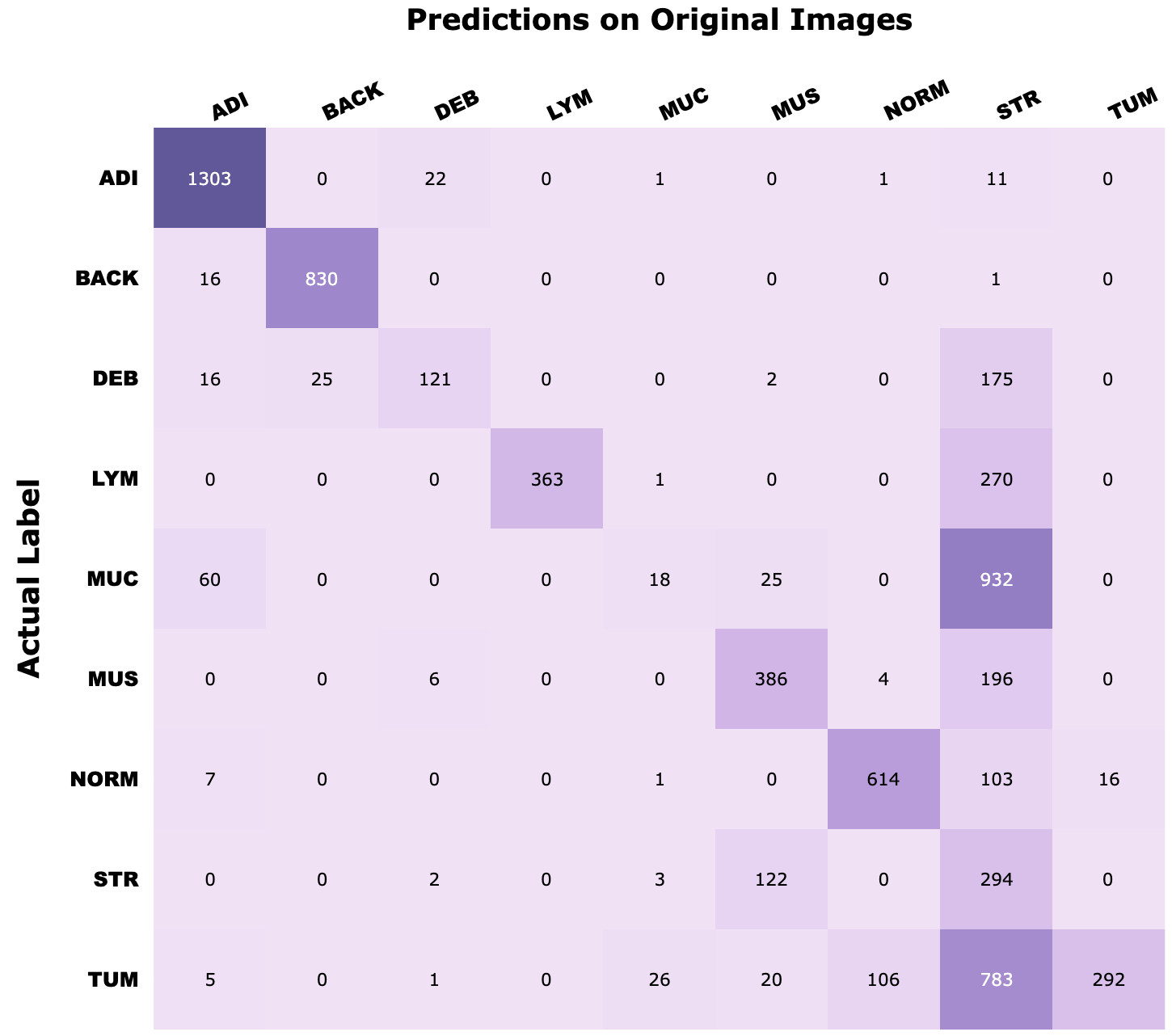}
\label{fig:heatmap_predictions_original}}\
{\includegraphics[width=0.4\textwidth]{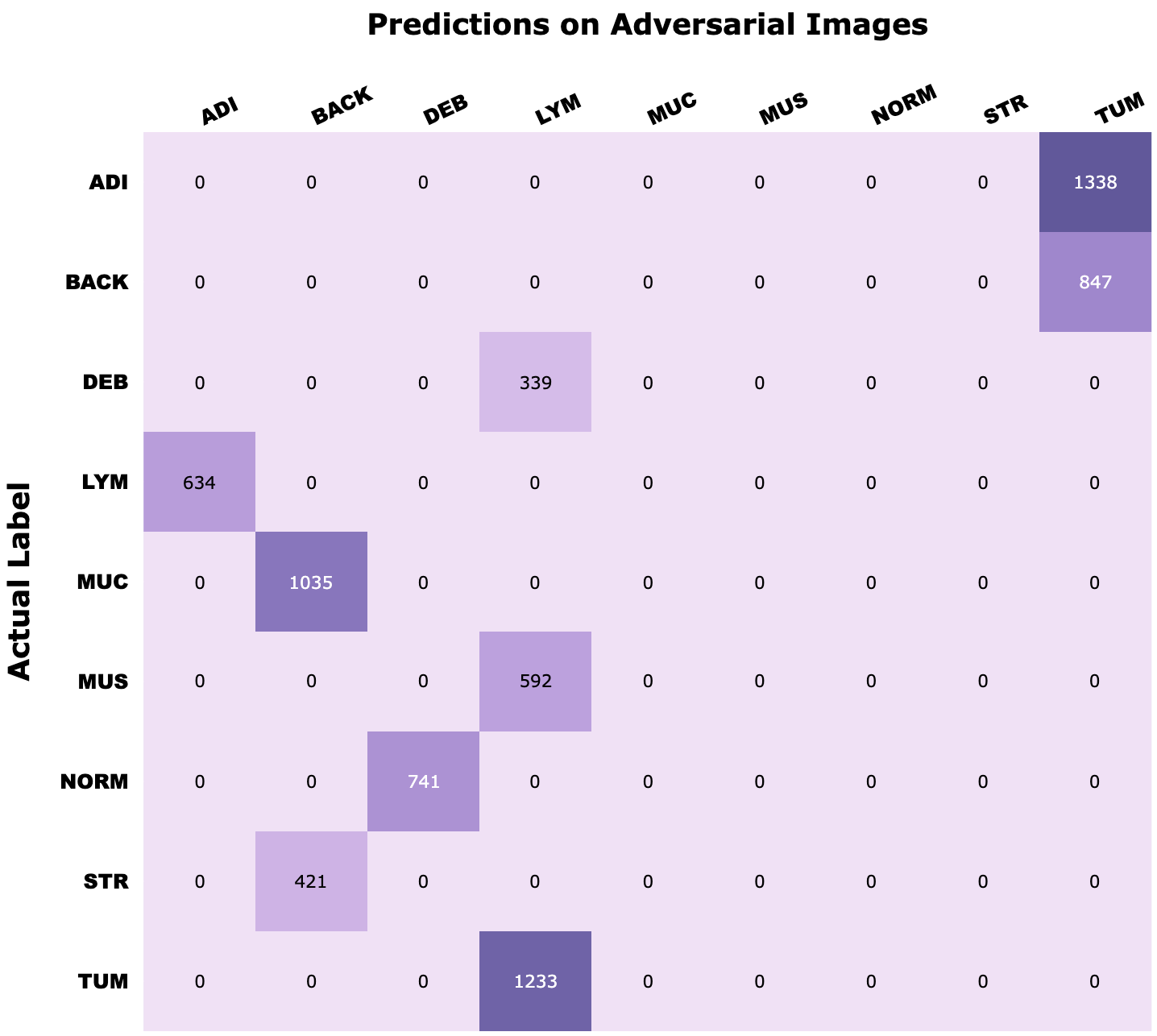}
\label{fig:heatmap_predictions_adversarial}}\
\caption{Heatmaps showing distribution of Adversarial Attacks}
\label{fig:5}
\end{figure}

In order to provide a clearer understanding of the influence of adversarial attacks on the predictions made by PLIP, we utilized heat maps as a visual tool to depict the model's label predictions both prior to and subsequent to the attacks, see Fig. \ref{fig:5}. A continuous pattern of accurate label predictions is observed in the top heat map obtained from the original photos. However, the observed pattern is significantly disturbed in the associated adversarial heat maps on the botton of Fig. \ref{fig:5}. This investigation highlights the difficulty presented by adversarial manipulations, placing emphasis on the significance of interpretability in medical artificial intelligence (AI) models.

\begin{figure}[htb]

\begin{minipage}[b]{1.0\linewidth}
  \centering
  \centerline{\includegraphics[width=8.5cm]{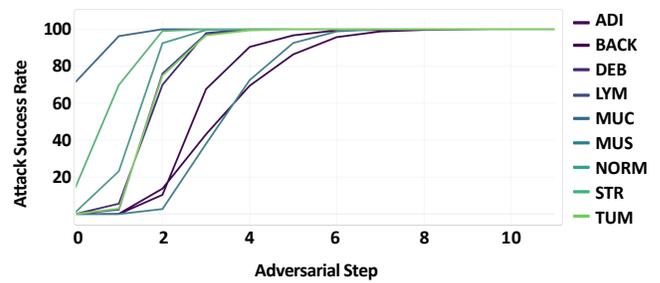}}
\end{minipage}
\caption{Attack Success Rate per step}
\label{fig:6}
\end{figure}

To strengthen our assessment, we calculated the SSIM ratings for both the unaltered and manipulated images.  Notably, it was noted that the majority of SSIM values are above 90\%, suggesting a significant level of resemblance between the unaltered and modified photos. The targeted PGD attacks aimed at inducing misclassifications within PLIP's predictions for specific tissue types were remarkably successful, yielding a 100\% attack success rate. From Fig. \ref{fig:6}, we observe that we have reached 100\% attack sccess rate for all the tissue types after 10 steps to achieve optimal pertubation, demonstrating the vulerability of the vision language model.

\begin{figure}[htb]
    \centering    \centerline{\includegraphics[width=1.0\linewidth]{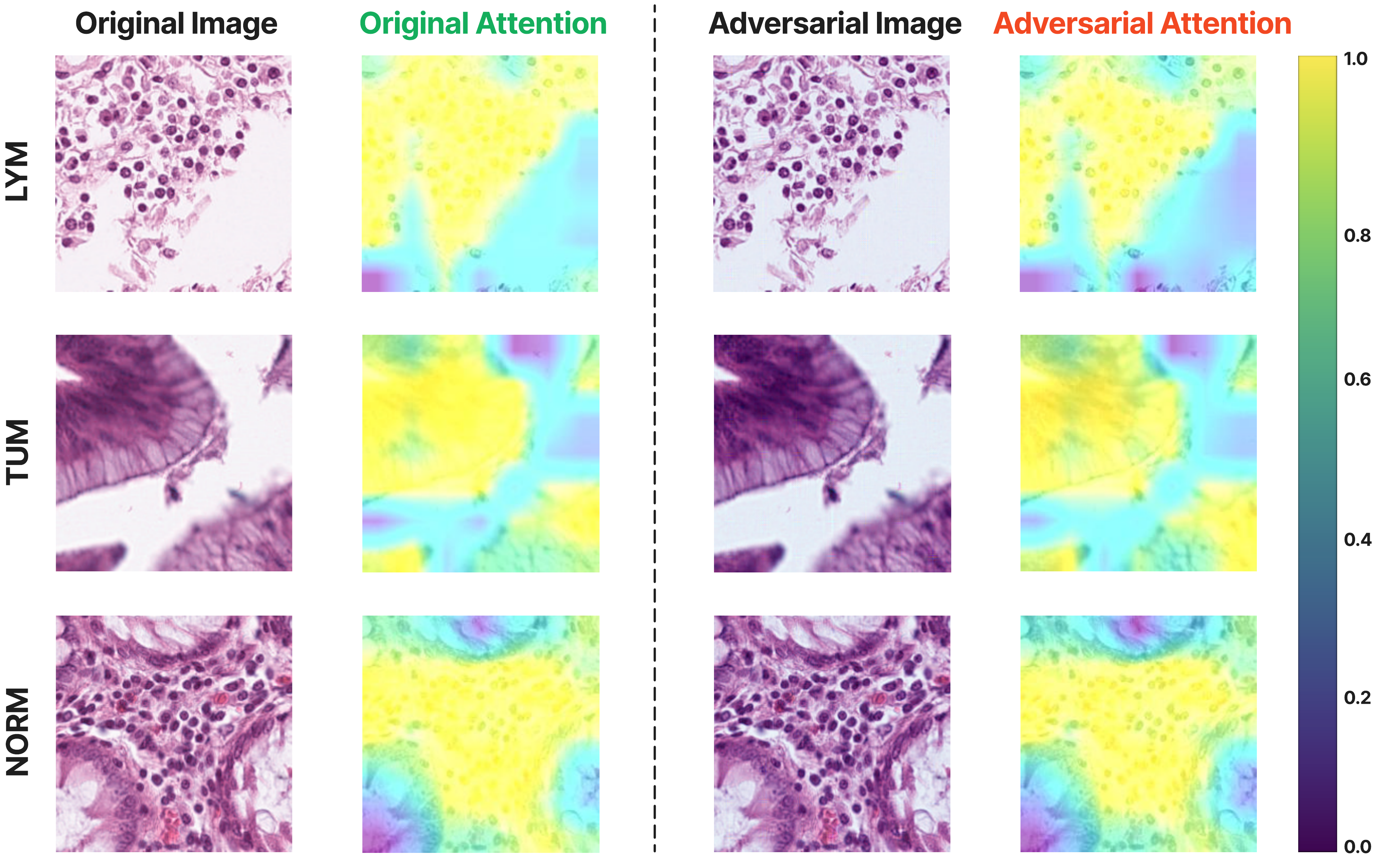}}
    \caption{Visualization of Attention before and after PGD attack on PLIP}
    \label{fig:4}
\end{figure}

Moreover, we visualized the attention patterns of the PLIP model before and after subjecting it to a PGD attack, i.e. both on the original and perturbed images. This examination provided valuable insights into the specific areas where the model focuses its attention and identifies regions crucial for classification \cite{aflalo2022vl}.

Upon inspecting the tissue images labeled LYM and NORM in Fig. \ref{fig:4}, we observed distinct attention patterns. In the original attention distribution (left side of Fig. \ref{fig:4}), the model exhibited heightened attention (indicated by the predominance of yellow) on various cell structures. However, following the PGD attack, represented by the adversarial attention distribution (right side of Fig. \ref{fig:4}), we noticed a noticeable shift or reduction in attention towards these cell structures. In particular, the attention allocated to these structures decreased (illustrated by the prevalence of green) compared to the original attention distribution. Interestingly, the attention appeared to redistribute towards the surrounding areas of the cell structures.

This analysis underscores the dynamic nature of the model's attention mechanism and highlights its susceptibility to adversarial perturbations. By discerning these changes in attention allocation, we gain valuable insights into the model's decision-making process and its sensitivity to external manipulations.

\subsection{Targeted Misclassification}
\label{ssec:subhead}

In the context of our targeted PGD adversarial attacks on the PLIP model using the Kather Colon dataset, our objective was to intentionally induce misclassifications for specific tissue types. 


These targeted misclassifications were chosen to assess the model's susceptibility to adversarial manipulations across a spectrum of tissue types, mirroring potential real-world scenarios, where misdiagnoses could have critical consequences. The successful implementation of these targeted misclassifications, which can be observed from Fig. \ref{fig:2}, further underscores the nuanced vulnerabilities of PLIP to adversarial attacks and raises questions about the model's trustworthiness in medical imaging applications. Along with PLIP, we have also performed this trageted PGD adversarial attack on an additional VLM model, BiomedClip, in which we have observed similar vulnerabilities of misclassifications (data not shown). \cite{zhang2024biomedclip} 

    \section{Discussions}
\label{sec:Discussions}

\subsection{Possible Defenses}
In the light of vulnerabilities exibhited by PLIP model under targeted PGD adversarial attack, incorporating robust defense strategies becomes necessary to safeguard model against such threats. 
To improve robustness, Adversarial training has been proved to be one of the most effective approaches in the image domain~\cite{xu2021robust}. 
However, understanding that the approach of adversarial training is computationally intensive, especially in the case of VLMs, recent works have explored input pre-processing techniques such as diffusion-based purification, to improve robustness of VLMs~\cite{nie2022diffusion}~\cite{shi2021online}.
These methods rely on generative models to purify adversarial perturbations before classification, without any need for re-training.
Given the critical nature of pathology image analysis and the potential implications of adversarial attacks in medical diagnostics, VLMs can be made more resilient to adversarial attacks by exploring a combination of these advanced strategies, thereby ensuring their security and reliability in real-world applications.

    \section{Conclusion and Future Directions}
\label{sec:conclusion}

In the pursuit of advancing the interpretability, domain adaptation, and trustworthiness of medical artificial intelligence, our investigation into the vulnerabilities of the PLIP model through targeted PGD adversarial attacks reveals critical insights. The 100\% success rate in inducing intentional misclassifications underscores PLIP's susceptibility to adversarial manipulations, prompting a fundamental reassessment of its trustworthiness in medical imaging applications. The qualitative analysis of adversarial examples illuminate the changes in PLIP's predictions, emphasizing the need for interpretability-aware defenses to fortify models. These findings contribute to the broader discourse on the robustness of Vision Language Models in pathology analysis, guiding the development of AI models that not only exhibit high performance but also maintain reliability in the face of adversarial attacks.


    \section{Acknowledgments}
\label{sec:acknowledgments}

 This work was supported by a University of Texas System Rising STARs Award (J.M.L) and the CPRIT First Time Faculty Award (J.M.L) 

 \section{Compliance with ethical standards}
This research study was conducted retrospectively using human subject data made available in open access. Ethical approval was  \textbf{not} required as confirmed by the license attached with the open access data.
    
    
    \bibliographystyle{IEEEbib}
    \bibliography{refs}

\begin{thebibliography}{10}

\bibitem{lu2021ai}
Ming~Y Lu, Tiffany~Y Chen, Drew~FK Williamson, Melissa Zhao, Maha Shady, Jana Lipkova, and Faisal Mahmood,
\newblock ``Ai-based pathology predicts origins for cancers of unknown primary,''
\newblock {\em Nature}, vol. 594, no. 7861, pp. 106--110, 2021.

\bibitem{ng2023prospective}
Annie~Y Ng, Cary~JG Oberije, {\'E}va Ambr{\'o}zay, Endre Szab{\'o}, Orsolya Serf{\H{o}}z{\H{o}}, Edit Karpati, Georgia Fox, Ben Glocker, Elizabeth~A Morris, G{\'a}bor Forrai, et~al.,
\newblock ``Prospective implementation of ai-assisted screen reading to improve early detection of breast cancer,''
\newblock {\em Nature Medicine}, pp. 1--6, 2023.

\bibitem{de2021deep}
Kevin de~Haan, Yijie Zhang, Jonathan~E Zuckerman, Tairan Liu, Anthony~E Sisk, Miguel~FP Diaz, Kuang-Yu Jen, Alexander Nobori, Sofia Liou, Sarah Zhang, et~al.,
\newblock ``Deep learning-based transformation of h\&e stained tissues into special stains,''
\newblock {\em Nature communications}, vol. 12, no. 1, pp. 4884, 2021.

\bibitem{Huang2023}
Z.~Huang, F.~Bianchi, M.~Yuksekgonul, T.~J. Montine, and J.~Zou,
\newblock ``A visual–language foundation model for pathology image analysis using medical twitter,''
\newblock {\em Nat Med}, vol. 29, no. 9, pp. Art. no. 9, Sep 2023.

\bibitem{goodfellow2014explaining}
Ian~J Goodfellow, Jonathon Shlens, and Christian Szegedy,
\newblock ``Explaining and harnessing adversarial examples,''
\newblock {\em arXiv preprint arXiv:1412.6572}, 2014.

\bibitem{carlini2017towards}
Nicholas Carlini and David Wagner,
\newblock ``Towards evaluating the robustness of neural networks,''
\newblock in {\em 2017 ieee symposium on security and privacy (sp)}. Ieee, 2017, pp. 39--57.

\bibitem{madry2019deep}
Aleksander Madry, Aleksandar Makelov, Ludwig Schmidt, Dimitris Tsipras, and Adrian Vladu,
\newblock ``Towards deep learning models resistant to adversarial attacks,'' 2019.

\bibitem{kather2019predicting}
Jakob~Nikolas Kather, Johannes Krisam, Pornpimol Charoentong, Tom Luedde, Esther Herpel, Cleo-Aron Weis, Timo Gaiser, Alexander Marx, Nektarios~A Valous, Dyke Ferber, et~al.,
\newblock ``Predicting survival from colorectal cancer histology slides using deep learning: A retrospective multicenter study,''
\newblock {\em PLoS medicine}, vol. 16, no. 1, pp. e1002730, 2019.

\bibitem{gamper2019pannuke}
Jevgenij Gamper, Navid Alemi~Koohbanani, Ksenija Benet, Ali Khuram, and Nasir Rajpoot,
\newblock ``Pannuke: an open pan-cancer histology dataset for nuclei instance segmentation and classification,''
\newblock in {\em Digital Pathology: 15th European Congress, ECDP 2019, Warwick, UK, April 10--13, 2019, Proceedings 15}. Springer, 2019, pp. 11--19.

\bibitem{da2022digestpath}
Qian Da, Xiaodi Huang, Zhongyu Li, Yanfei Zuo, Chenbin Zhang, Jingxin Liu, Wen Chen, Jiahui Li, Dou Xu, Zhiqiang Hu, et~al.,
\newblock ``Digestpath: A benchmark dataset with challenge review for the pathological detection and segmentation of digestive-system,''
\newblock {\em Medical Image Analysis}, vol. 80, pp. 102485, 2022.

\bibitem{aflalo2022vl}
Estelle Aflalo, Meng Du, Shao-Yen Tseng, Yongfei Liu, Chenfei Wu, Nan Duan, and Vasudev Lal,
\newblock ``Vl-interpret: An interactive visualization tool for interpreting vision-language transformers,''
\newblock in {\em Proceedings of the IEEE/CVF Conference on Computer Vision and Pattern Recognition}, 2022, pp. 21406--21415.

\bibitem{zhang2024biomedclip}
Sheng Zhang, Yanbo Xu, Naoto Usuyama, Hanwen Xu, Jaspreet Bagga, Robert Tinn, Sam Preston, Rajesh Rao, Mu~Wei, Naveen Valluri, Cliff Wong, Andrea Tupini, Yu~Wang, Matt Mazzola, Swadheen Shukla, Lars Liden, Jianfeng Gao, Matthew~P. Lungren, Tristan Naumann, Sheng Wang, and Hoifung Poon,
\newblock ``Biomedclip: a multimodal biomedical foundation model pretrained from fifteen million scientific image-text pairs,'' 2024.

\bibitem{xu2021robust}
Han Xu, Xiaorui Liu, Yaxin Li, Anil Jain, and Jiliang Tang,
\newblock ``To be robust or to be fair: Towards fairness in adversarial training,''
\newblock in {\em International conference on machine learning}. PMLR, 2021, pp. 11492--11501.

\bibitem{nie2022diffusion}
Weili Nie, Brandon Guo, Yujia Huang, Chaowei Xiao, Arash Vahdat, and Anima Anandkumar,
\newblock ``Diffusion models for adversarial purification,''
\newblock {\em arXiv preprint arXiv:2205.07460}, 2022.

\bibitem{shi2021online}
Changhao Shi, Chester Holtz, and Gal Mishne,
\newblock ``Online adversarial purification based on self-supervised learning,''
\newblock in {\em International Conference on Learning Representations}, 2021.

\end{thebibliography}

\end{document}